\documentclass[
reprint,
groupedaddress,
 amsmath,amssymb,
 aps,
prb,
]{revtex4-2}

\usepackage{dcolumn}
\usepackage{graphicx}
\usepackage{hyperref}
\usepackage{color}
\usepackage{tikz}
\usepackage{ascmac,amsmath,amsfonts,amsthm,amssymb}
\usepackage{url}
\usepackage{nicematrix}

\usepackage[whole]{bxcjkjatype}

\newenvironment{itmbx}[1]{\begin{itembox}[l]{#1}}{\end{itembox}}

\newcommand{\tr}{\mathrm{Tr}}     

\newcommand{\ket}[1]{| #1 \rangle}
\newcommand{\bra}[1]{\langle #1 |}

\begin{document}

\preprint{APS/123-QED}

\title{Gap labeling theorem for multilayer thin film heterostructures}

\author{Mao~Yoshii}
\email{mao@g.ecc.u-tokyo.ac.jp}
\affiliation{Department of Applied Physics, The University of Tokyo, Hongo, Tokyo, 113-8656, Japan}
\author{Sota~Kitamura}
\affiliation{Department of Applied Physics, The University of Tokyo, Hongo, Tokyo, 113-8656, Japan}
\author{Takahiro~Morimoto}
\affiliation{Department of Applied Physics, The University of Tokyo, Hongo, Tokyo, 113-8656, Japan}

\date{\today}

\begin{abstract}
Quasiperiodic systems show a universal  gap structure due to quasiperiodicity which is analogous to gap openings at the Brillouin zone boundary in periodic systems. The integrated density of states (IDoS) below those energy gaps are characterized by a few integers, which is known as the ``gap labeling theorem'' (GLT) for quasiperiodic systems.
In this study, focusing on multilayer thin film systems such as twisted bilayer graphene and stacked transition metal dichalcogenides, we extend the GLT for multilayer systems of arbitrary dimensions and number of layers, using an approach based on the algebra called ``a noncommutative torus''.
We find that the energy gaps and the associated IDoS are generally characterized by $_{DN}C_D$ integer labels in $N$ layer systems in the $D$ dimensions, when the effect of the interlayer coupling can be approximated by a quasiperiodic intralayer coupling for each layer. We demonstrate that the generalized GLT holds for quasiperiodic 1D tight binding models by numerical simulations.
\end{abstract}

\maketitle

\section{Introduction}
Quasiperiodic systems are systems that possess long-range order without translational symmetry. In 1982, quasiperiodic structure is discovered in the system of alloys~\cite{PhysRevLett.53.1951}, and quasiperiodicity has later been found in various systems \cite{PhysRevLett.109.116404,vardeny2013optics,kamiya2018discovery,tsai2000stable,collins2017imaging}. The structure of quasiperiodic crystals can be regarded as a projection of higher-dimensional-crystalline structure \cite{DEBRUIJN198139,DEBRUIJN198153}, and would allow us to access the physics of higher-dimensional-space that is usually inaccessible in three-dimensional crystals. 
Recently, stacked system of two-dimensional thin films has been realized and intensively studied, including twisted bilayer graphenes  \cite{Bistritzer12233,cao2018correlated,PhysRevB.99.165430} and interface of transition metal dichalcogenides \cite{wang2020correlated,Akamatsu68}.
Multilayer systems made of different crystals can be also considered as quasiperiodic systems \cite{Akamatsu68,kennes2021moire}, which provides an interesting platform for quasiperiodic structures due to their controllability and a rich variety of material combinations.  

In periodic systems, the energy gap often opens at the Brillouin zone (BZ) boundary due to anticrossing of energy bands that are related by the reciprocal vectors.
Similarly, in quasiperiodic systems, there exist energy gaps that originate from quasiperiodicity.
In the quasiperiodic systems, BZ folding takes place in the momentum space picture and leads to replicas of energy bands that exhibit anticrossings.
Since those energy gaps stem from the geometry of the quasiperiodic systems, one can relate the energy gaps with the geometric parameters of the system.
Since the energy structure of quasiperiodic systems cannot be captured by the energy dispersion in the momentum space picture in general, understanding the energy gap structure independent of the system size are particularly important.
For example, it is known that gaps in the energy spectrum of Fibonacci quasicrystal can be labeled by two integers~\cite{PhysRevB.93.205153,BENAMEUR2002667,PhysRevB.46.9216,PhysRevB.66.094202}.
Using those gap labels, one can discuss the physical property of the energy gaps regardless of the system size.

Multilayer thin films (MLTFs) show a universal energy gap structure which comes from the quasiperiodicity.
Such gap structure can be understood using the so called ``gap labeling theorem'' (GLT).
The GLT establishes a relationship between those energy gaps with the integrated density of states (IDoS) below the gap through integer labels, and is known for quasicrystals for several decades \cite{PhysRevB.46.9216,PhysRevB.66.094202}.
A GLT for MLTFs was originally proposed using an algebra called noncommutative torus (NCT) in the case of two-dimensional homo-bilayer systems (which are made of the same type of atomic layers)~\cite{rosa2021topological}.
In Ref.~\cite{rosa2021topological}, the GLT for a twisted bilayer system consisting of the same type of atomic layers was derived, which states that the IDoS below a certain energy gap $G$ is given by
\begin{align}\label{eq : IDoS homo-bilayer intro}
    \mathrm{IDoS}(G) =& \mathfrak{n}_{\emptyset} + \sum_{i,j=1,2}\mathfrak{n}_{ij}\frac{| \boldsymbol{a}_{i} \times \boldsymbol{b}_{j} |}{S}
    \quad
    (\mathfrak{n}_{ij} \in \mathbb{Z}).
\end{align}
Here, $\boldsymbol{a}_i$ is a primitive vector of one layer, $\boldsymbol{b}_j$ is a primitive vector of the other layer, and $S$ is $|\boldsymbol{a}_1\times \boldsymbol{a}_2|$.
Namely, the GLT gives a labeling for the IDoS below the energy gap $G$ with integers $\mathfrak{n}$.

\begin{figure*}
    \centering
    \includegraphics[width = \linewidth]{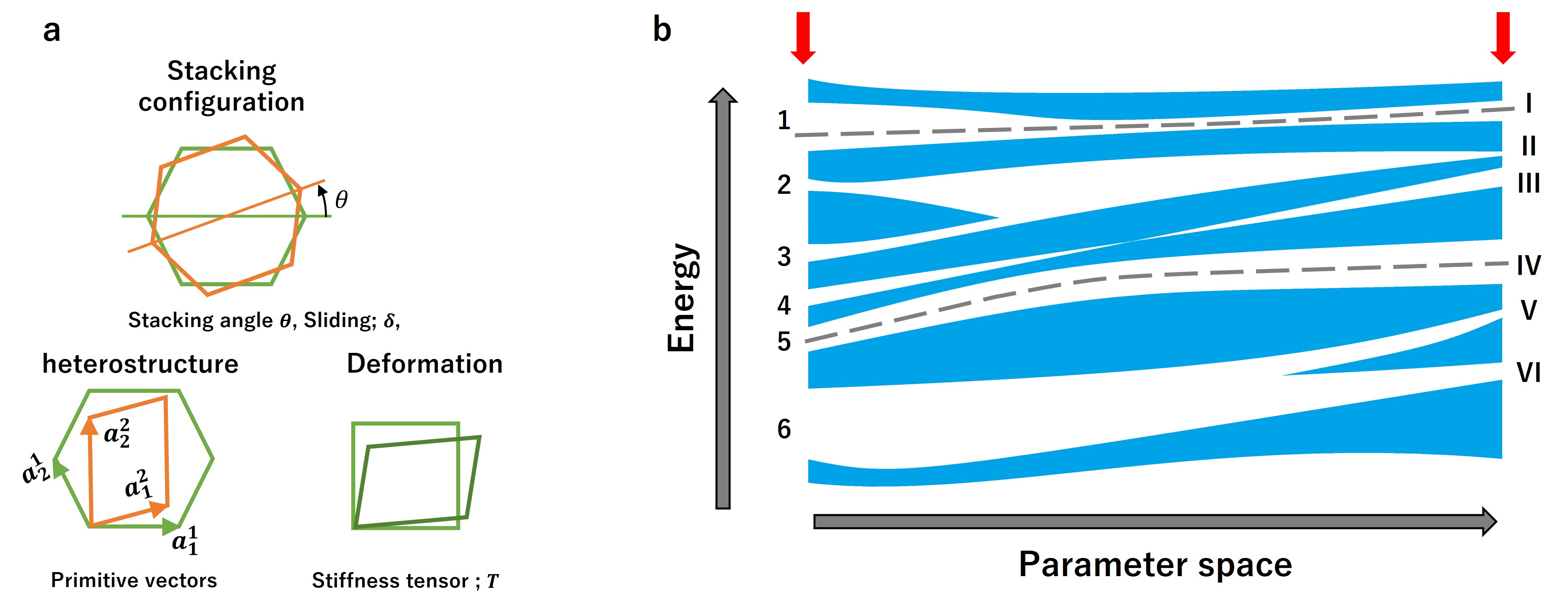}
    \caption{
    (a) Parameters that characterize the MLTFs. They include stacking angle, lattice constants of the stacked layers, and deformation strength.
    (b) A schematic picture of the energy spectrum of MLTFs when one changes the parameters characterizing the MLTFs.
    Blue regions represent the energy bands and the white regions the energy gaps.
    Gray dashed lines are the energy gaps that originates from the quasiperiodicity and can be captured by the GLT.
    Those energy gaps show a robust behavior, while other nonuniversal gaps originating from band hybridizations can merge or split when the parameters are changed.
    }
    \label{fig : Energy band and parameter space}
\end{figure*}

The GLT is also useful to understand the behavior of the energy gaps when one changes a parameter that characterizes the quasiperiodic system.
For example, if we consider a twisted bilayer system, $|\boldsymbol{a}_{i} \times \boldsymbol{b}_{j} |$ changes continuously with varying the twist angle (Fig.~\ref{fig : Energy band and parameter space}a).
In this case, the energy gaps for two twist angles can be smoothly connected with each other when the corresponding IDoS have the same label $\mathfrak{n}$, as depicted by the dashed curve in Fig.~\ref{fig : Energy band and parameter space}b. 
More recently, it has been pointed out that such energy gap structure can be also understood from charge transport that appears when sliding two thin films relatively, which is characterized by a topological quantity called the sliding Chern number (SCN) \cite{PhysRevB.101.041112,PhysRevResearch.4.013028}.

Since derivation of the GLT from the NCT approach only requires information of the quasiperiodic lattice structure, the NCT approach is concise and systematic, compared to the SCN approach.
So far, the NCT approach for the GLT was only applied for two-dimensional homo-bilayer systems \cite{rosa2021topological}. 
Hence, it is an interesting problem to extend the NCT method to MLTF heterostructures made of different types of 2D systems.
An obstacle in extending the NCT approach for MLTF heterostructures is that the information of the relative size of the unit cells of different layers are not incorporated in the original NCT construction, while it is indispensable to derive GLT for multilayer thin films.
Also, previous studies on homo-bilayer systems~\cite{rosa2021topological,PhysRevB.101.041112,PhysRevResearch.4.013028} derived GLT by focusing on the IDoS of one layer, since the system is invariant under exchanging the two layers in homo-bilayer systems,
which is no longer the case for general MLTF heterostructures.
Therefore, to understand the electronic structure of general MLTFs, it is necessary to treat the electronic structure of all layers on an equal footing.

In this paper, we extend the NCT approach for the GLT to MLTF heterostructures.
To this end, we adopt an approach from the approximately finite $C^*$-algebra \cite{effros1980approximately,10.2307/24714004}. 
Specifically, we relate the traces for the different layers using the inductive limit for the parameters characterizing the NCT obtained from the continued fraction,
which is effectively equivalent to considering large approximant of MLTF heterostructures.
Assuming that the interlayer coupling is approximated by quasiperiodic intralayer couplings for each layer, which generally holds when states in different layers are energetically separated,
we extend the GLT for MLTF heterostructures.
The generalized GLT reveals that the IDoS for $N$ layer heterostructures in the $D$ dimensions is characterized by $_{DN}C_D$ integer labels.
We perform numerical simulations for tight binding models of 1D MLTF heterostructures and demonstrate that the generalized GLT indeed holds.

The rest of the paper is organized as follows.
In Sec.~\ref{sec : Gap labelling theorem}, we review the GLT for MLTFs consisting of the same type of atomic layers, and introduce algebraic structure of operators in the MLTFs.
In Sec.~\ref{sec : Gap labelling theorem of one-dimensional multilayer thin films}, we first extend the GLT to the bilayer heterostructures in one-dimension and then to the general $N$ layer systems in arbitrary dimensions in Sec.~\ref{sec : Gap labelling theorem of $D$-dimensional MLTFs}.
In Sec.~\ref{sec : Numerical calculation}, we show numerical calculation in one-dimensional MLTF heterostructure system to demonstrate that the generalized GLT holds.

\section{Gap labeling theorem \label{sec : Gap labelling theorem}}

In this section, we briefly introduce the original gap labeling theorem \cite{BENAMEUR2002667,PhysRevB.66.094202,PhysRevB.46.9216}.
In periodic systems, band folding at the Brillouin zone boundary often leads to anticrossing between folded bands and gap opening. When the energy gap appears from such band folding, the integrated density of states (IDoS) below such gap is given by an integer times BZ volume.
In quasiperiodic systems, a gap opening also appears from the band folding due to the quasiperiodic structure and a similar relationship for IDoS holds, which is known as GLT.

To explain the GLT, we first define the IDoS as follows. 
Let $\hat{H}$ a Hamiltonian with energy eigenvalues $E_1 \leq \cdots \leq E_{\dim H}$ and eigenstates $\{\ket{\psi_n}\},~(n \in \{ 1, \ldots, \dim H \} )$.
We denote projector to the states below the energy $E$ as
\begin{align}
    \hat{P}(E) = \sum_{E_i \leq E} \ket{\psi_i}\bra{\psi_i} .
\end{align}
Since we mainly concentrate on the energy gaps in this paper, it is convenient to relate $\hat{P}(E)$ to energy gaps.
When we name the energy gap between $E_n$ and $E_{n+1}$ to be $G$, we also write $\hat{P}(G) := \hat{P}(E_n)$.
Using projector, IDoS is defined as
\begin{align}\label{eq : def tau}
    \mathrm{IDoS}(E) =& \tau(\hat{P}(E)), \\
    \tau(\hat{\mathcal{O}}) =& \frac{1}{\dim(\hat{\mathcal{O}})}\tr(\hat{\mathcal{O}}).
\end{align}
Here, $\tau$ is the normalized trace defined so that $\tau$ of identity operator $\boldsymbol{1}$ becomes $1$. 
In the same manner, we also write it as $\mathrm{IDoS}(G) = \tau(\hat{P}(G)).$
While we can define the normalized trace as in Eq.~\eqref{eq : def tau} for finite dimensions, in the case of infinite dimensions, we adopt the inductive limit from finite-dimensional matrix algebras for construction of NCT and definition of $\tau$ [with Eq.~(\ref{eq : def tau})]~\cite{gracia2000elements,Takesaki2003,kenneth1996c}.
In the rest of this section, to explain the GLT, we explicitly derive the GLT for a toy model.

\subsection{Gap labeling theorem in one dimension}
To extend the formalism of Ref.~\cite{rosa2021topological}, let us review the original GLT
in the case of one-dimension using the so-called Harper model.

Firstly, we demonstrate that the Hamiltonian of a lattice under an incommensurate potential can be regarded as an operator on NC torus.
Let us consider the following model,
\begin{align}\label{eq : H toy model}
    \hat{H} = \sum_{n} 
    \big[
    \left( t\ket{n+1}\bra{n} + h.c. \right) 
    +
    V(n)\ket{n}\bra{n}
    \big],
\end{align}
where $V(x) = V(x+\theta^{-1})$ is a periodic function, which is incommensurate when $\theta$ is taken to be irrational.
We can expand $V$ into a Fourier series as
\begin{align}
    V(x) = \sum_{\eta \in \mathbb{Z}} v_{\eta} e^{2\pi i \eta x \theta}.
\end{align}
Let us define $\hat{S}$ and $\hat{U}$ as
\begin{align}
    \hat{S} =& \sum_{n}\ket{n-1}\bra{n}, \\
    \hat{U} =& \sum_{n} e^{2\pi i n\theta} \ket{n}\bra{n},
\end{align}
which follow the commutation relation
\begin{align}
    \hat{S}\hat{U} = e^{2\pi i \theta} \hat{U}\hat{S}.
\end{align}
With these operators, we can write the Hamiltonian as a polynomial of $\hat{U}$ and $\hat{S}$ as
\begin{align}
    \hat{H} = t(\hat{S} + \hat{S}^\dagger) + \sum_{\eta \in \mathbb{Z}} v_{\eta}\hat{U}^{\eta}.
\end{align}
\subsection{Noncommutative torus}
The above commutation relation defines an algebra called noncommutative torus.
The noncommutative $D$-torus is an algebra of operators $\{\hat{U}_j\}_{j = 1,\ldots,D}$ which follows the commutation relation,
\begin{align}
    \hat{U}_i\hat{U}_j = e^{2\pi i \theta_{ij}} \hat{U}_j\hat{U}_i.
\end{align}
Here, $\theta_{ij}$ is a real number which characterizes noncommutative torus,
which can be regarded as an element of the anti-symmetric matrix $\theta$.
We call $\theta$ as a noncommutative parameter and denote noncommutative $D$-torus defined by $\theta$ as $\mathbb{T}^D_{\theta}$.
For example, a noncommutative two-torus appears in the context of Hofstadter's butterfly in quantum Hall systems.
In noncommutative $D$-torus, we can also construct a projection operator, whose trace is directly related to the IDoS. 
For any Hamiltonian in an algebra of the noncommutative $D$-torus $\mathbb{T}^D_{\theta}$, it is known that the IDoS of the state below energy gap $G$ is expressed using $\theta$ as \cite{prodan2016bulk}
\begin{align}
    \mathrm{IDoS}(G) = \mathfrak{n}_{\emptyset} + \sum_{J \subset \{1,\ldots , D\}} \mathfrak{n}_J \mathrm{Pf}(\theta_{J}),
    \label{eq: IDos G}
\end{align}
where the subscript $J$ labels subsets of $\{1,\ldots , D\}$, $\mathrm{Pf}$ denotes a Pfaffian and $\theta_{J}$ is a submatrix of $\theta$ defined as 
\begin{align}\label{eq : def theta n}
    \{\theta_{J} \}_{ij} = 
    \begin{cases}
        \theta_{ij} & (i,j \in J) \\
        0 & \mathrm{others}
    \end{cases}.
\end{align}
Once we obtain the parameter $\theta$ for the quasiperiodic system, application of the above formulae leads to the expression for the IDoS.

For the case of Eq.~(\ref{eq : H toy model}), $\hat{H}$ is an operator belonging to the NCT of parameter $\theta$, the associated IDoS is given by
\begin{align}\label{eq : IDoS of NCT}
    \mathrm{IDoS}(G) = \mathfrak{n}_{\emptyset} + \theta \mathfrak{n}_{1}.
\end{align}

\section{Gap labeling theorem of one-dimensional multilayer thin films}\label{sec : Gap labelling theorem of one-dimensional multilayer thin films}

In this section, we focus on a one-dimensional system and derive the generalized GLT for MLTF heterostructures using the NCT approach,
which is consistent with those obtained from other approaches \cite{PRODAN2019135,PhysRevB.101.041112}.

One obstacle in extending the GLT is the relation of the normalized trace $\tau$ in MLTF systems.
As we show below, we compute the gap labels of the MLTF heterostructres 
by approximately decoupling layers and applying NCT approach to each layer.
However, since the normalized trace $\tau$ has a different normalization factor for a different Hilbert space,
the IDoS of the entire MLTF is not simply obtained as a sum of the IDoS of each layer.
To derive the gap label for the MLTFs, we need to know the relation between the $\tau$s defined in each layer.
Specifically, for the infinite-dimensional case,
we take inductive limit with finite-dimensional matrix algebras, where
the relation of the normalized trace for different layers is incorporated in a straightforward way.
This approach is known as the approximately finite (AF) algebra \cite{effros1980approximately,10.2307/24714004,gracia2000elements}.

In this section, we focus on one-dimensional MLTF heterostructures and
apply the above operations to derive the generalized GLT.
We mostly consider the bilayer case, and mention the general $N$ layer systems in the end.

\subsection{Hamiltonian for a quasiperiodic bilayer}
We construct a Hamiltonian $\hat{H}$ for a quasiperiodic bilayer by coupling two single-layer tight-binding models (we denote the single-layer model of layer $l$ as $\hat{H}^{l}$) with interlayer coupling $\hat{W}$,
\begin{align}\label{eq : Hamiltonian 1D 2L}
    \hat{H} = \hat{H}^1 + \hat{H}^2 + \hat{W} + \hat{W}^\dagger.
\end{align}
Here, the single-layer Hamiltonian $\hat{H}^l$ does not depend on the other layers, and has no internal degrees of freedom (such as sublattice degrees of freedom) for simplicity. Namely,
\begin{align}
    \hat{H}^l = \sum_{n^l,m^l\in\mathbb{Z}} h_{n^l-m^l}\ket{n^l,l}\bra{m^l,l}.
\end{align}
Here, $n^l$ and $m^l$ are the label of sites on layer $l$.
Hereafter we omit the second (layer) index $l$ for states as it is apparent from the superscript of the first (site) index $n^l$.
The position of site $n^l$ is defined as $n^la^l$ with $a^l$ being the lattice constant of layer $l$.
From the periodicity of layer $l$, the hopping amplitude $h$ depends only on the relative distance between site $n^{l}$ and site $m^{l}$.
In contrast, the interlayer coupling $\hat{W}$ depends on how the two layers are stacked.
We express $\hat{W}$ by a fully connected model as 
\begin{align}\label{eq : def interlayer coupling}
    \hat{W} =& \sum_{n^1,m^2}
    W(n^1,m^2) \ket{n^1} \bra{m^2}.
\end{align}

\subsection{Effective Hamiltonian in each layer}\label{sec : Effective Hamiltonian in each layer}
The GLT from the NCT approach can capture a gap structure that arises from the quasiperiodicity, which we call intraband energy gap. 
Such energy gap originates from gap opening for folded bands due to lattice constant mismatch and appears at the Bragg lines, which is analogous to the gap opening at the BZ boundary in periodic systems. 
On the other hand, interlayer coupling in MLTFs also causes hybridization between different bands, and leads to gap opening that does not directly related to the quasiperiodic nature, which we call interband energy gap.
In order to capture the universal gap structure of intraband energy gaps by the NCT approach, below we approximate the interlayer couplings in Eq.~(\ref{eq : Hamiltonian 1D 2L}) with quasiperiodic intralayer couplings.
This approximation is justified as far as energy bands of different layers are energetically separated and the interlayer coupling is not too large.

\begin{figure*}
    \centering
    \includegraphics[ width = \linewidth]{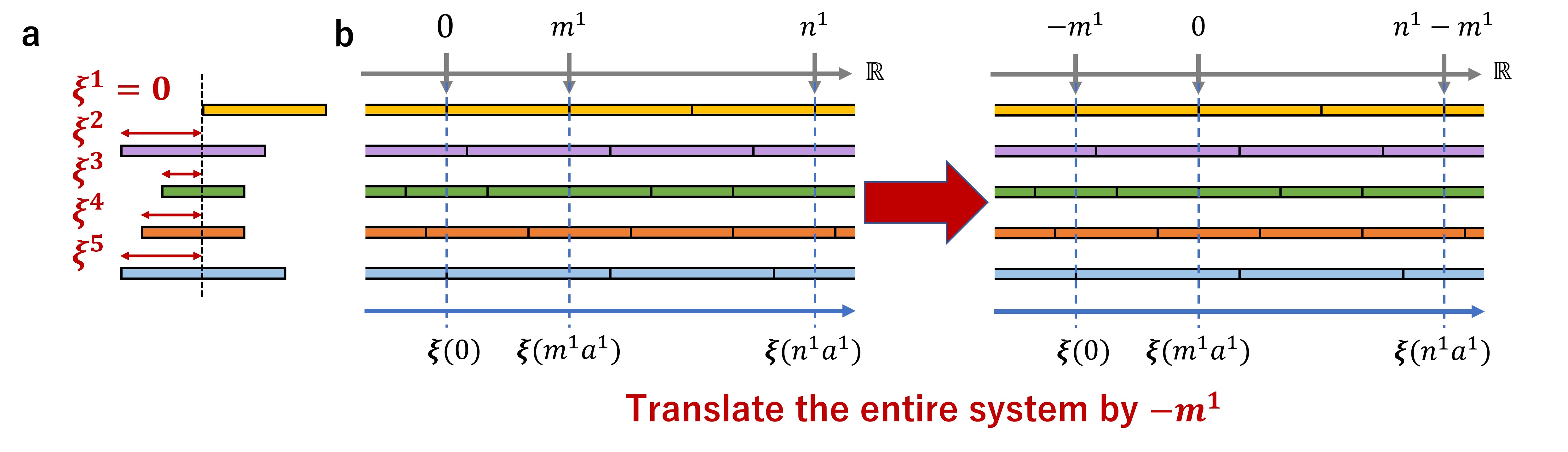}
    \caption{
    A schematic picture of the parameter $\xi$ characterizing the layer stacking of the MLTFs.
    (a) Example of $\boldsymbol{\xi}(\boldsymbol{r})$ of an electron on layer $1$.
    We focus on the a particular unit cell in the layer 1 and suppose that the electron is located at the position $\xi^1$. ($\xi^1=0$ corresponds to the base point of the unit cell in the layer 1.)
    For the layer $l$ other than 1, we define parameter $\xi^l$ by a position of the same electron measured within the unit cell of the layer $l$.
    In the tight-binding model, electrons are bound to the lattice sites and $\xi^1 = 0$.
    (b) Translation of the layers and $\boldsymbol{\xi}$.
    When we translate the layers by $-m^1a^1$, the original parameter $\boldsymbol{\xi}(0)$ is replaced by $\boldsymbol{\xi}(m^1a^1)$.
    }
    \label{fig : xi}
\end{figure*}

When we define a projector onto the space in layer $l$ as $\hat{P}^l$, we can write the equation $\hat{H}\ket{\psi} = E\ket{\psi}$ as
\begin{align}
    E \hat{P}^1 \ket{\psi} =& \hat{H}^1 \hat{P}^1\ket{\psi} + \hat{W} \hat{P}^2 \ket{\psi},
    \\
    E \hat{P}^2 \ket{\psi} =& \hat{W}^\dagger \hat{P}^1\ket{\psi} + \hat{H}^2 \hat{P}^2 \ket{\psi}.
\end{align}
and the effective Hamiltonian in layer $l$ is
\begin{align}
    \hat{H}^{1}_\text{eff}(E) = &
    \hat{H}^1
    +
    \hat{W} \frac{1}{E-\hat{H}^2} \hat{W}^\dagger,
    \label{eq : effective Hamiltonian of layer 1}
    \\
    \hat{H}^{2}_\text{eff}(E) = &
    \hat{H}^2
    +
    \hat{W}^\dagger \frac{1}{E-\hat{H}^1} \hat{W}.
    \label{eq : effective Hamiltonian of layer 2}
\end{align}
In the following, we approximate $E$ of $H^{1}_\text{eff} (H^{2}_\text{eff})$ with some constant which is comparable with eigenvalues of $H^1(H^2)$ of interest.
The second term of each line is the quasiperiodic term that behaves as an intraband coupling subject to the quasiperiodic pattern.
We denote such quasiperiodic intralayer coupling in layer $l$ as $\hat{V}^l$, which can be formally expressed as
\begin{align}
    \hat{V}^l = \sum_{n^l,m^l} \tilde{V}^{l}(n^l,m^l, \boldsymbol{\xi}(0))\ket{n^l} \bra{m^l}.
\end{align}
Here, $\xi^l$ is a function of $r \in \mathbb{R}$ which specifies the position in the unit cell of layer $l$ \cite{rosa2021topological}.
As depicted in Fig.~\ref{fig : xi}a, an electron at one layer defines the positions in unit cells of other layers $\xi^l$.
For a general position $r$ for an electron, collecting the position $\xi^l(r)$ in the unit cell of each layer, we write 
$\boldsymbol{\xi}(r) = \{ \xi^1(r), \ldots, \xi^N(r) \}$, 
which determines the stacking  configuration of the unit cells of different layers at the position $r$.
From the periodicity of layers, specifying how we stack unit cells is equivalent to specifying how we stack layers.
Thus, $\boldsymbol{\xi}$ at some point $r$ specifies the quasiperiodic pattern of the MLTF heterostucture. 
We mainly use $\boldsymbol{\xi}(0)$ to specify the quasiperiodic pattern in the following sections.

Next, we consider a term $\tilde{V}^{l}(n^l,m^l,\boldsymbol{\xi}(0))$.
This represents the hopping process where electron at site $m^l$ hops to $n^l$ in the MLTFs specified by $\boldsymbol{\xi}(0)$.
When we translate the system by $-m^la^l$, site $n^l$ is translated to $n^l-m^l$, and site $m^l$ is at site $0$.
Regarding $\boldsymbol{\xi}$, the $\boldsymbol{\xi}(0)$ of the translated system corresponds to $\boldsymbol{\xi}(m^la^l)$ in the original system (Fig.~\ref{fig : xi}b).
Hence, $\tilde{V}^{l}(n^l,m^l,\boldsymbol{\xi}(0))$ coincides with $\tilde{V}^{l}(n^l-m^l,0,\boldsymbol{\xi}(m^la^l))$ after the translation.
Therefore, omitting $0$ in the second argument, we can write $\hat{V}^l$ as
\begin{align}
    \hat{V}^l = \sum_{n^l,m^l} \tilde{V}^{l}(n^l-m^l, \boldsymbol{\xi}(m^la^l))\ket{n^l} \bra{m^l}.
\end{align}
When we consider layer $l$ under the tight-binding approximation, electrons are bound to the sites with no internal degrees of freedom.
In this case, $\xi^l(m^la^l) = 0$, and we may omit $\xi^l$ from $\boldsymbol{\xi}$.
\subsection{Fourier expansion}\label{sec : Fourier expansion}
Next, we expand $ \hat{V}^l$ in a Fourier series.
First, we consider the layer $l=1$ and define $q^1 = n^1-m^1$.
Applying the discussions in Sec.~\ref{sec : Effective Hamiltonian in each layer} to a bilayer system,
 $\xi^2$ has the periodicity $\xi^2(m^1a^1+a^2) = \xi^2(m^1a^1)$.
Using this periodicity, we expand the quasiperiodic intralayer coupling in the Fourier series as
\begin{align}
    V^1(q^1,\xi^2(m^1a^1))
    =
    \sum_{\eta^1 \in \mathbb{Z}}
    v_{q^1,\eta^1} e^{2\pi i \eta^1\frac{\xi^2(m^1a^1)}{a^2}}.
\end{align}
Defining the slide operator as $\hat{S}^1= \sum_{n}\ket{n^1-1}\bra{n^1}$, we can write
\begin{align}
    \hat{V}^1
    =& \sum_{m^1,q^1} \sum_{\eta^1}
    v_{q^1,\eta^1}
    e^{2\pi i \eta^1 \frac{m^1a^1}{a^2}}
    (\hat{S}^{1\dagger})^{q^1} \ket{m^1} \bra{m^1}
    +h.c.
\end{align}
where we have used $e^{2\pi i\frac{\xi^2(m^1a^1)}{a^2}} = e^{2\pi i \frac{m^1a^1}{a^2}}$.
Defining the generator $\hat{U}_1$ as
$\sum_{m^1} e^{2\pi i \frac{m^1a^1}{a^2}}\ket{m^1} \bra{m^1}$, we rewrite the above equation as
\begin{align}
    \hat{V}^1
    =& \sum_{q^1}\sum_{\eta^1} 
    v_{q^1,\eta^1}
    (\hat{S}^{1\dagger})^{q^1} 
    \hat{U}_1^{\eta^1} + h.c.
\end{align}
In this manner, we can express quasiperiodic intralayer coupling as a polynomial of $\hat{U}_1$ and $\hat{S}^1$.
In the same procedure, we can also decompose quasiperiodic intralayer coupling in layer $2$ as
\begin{align}
    \hat{V}^2
    =& \sum_{q^2}\sum_{\eta^2}
    v_{q^2,\eta^2}
    (\hat{S}^{2\dagger})^{q^2} 
    \hat{U}_2^{\eta^2}
    +h.c,
    \\
    \hat{U}_2 =& \sum_{m^2} e^{2\pi i \frac{m^2a^2}{a^1}}\ket{m^2} \bra{m^2},
    \\
    \hat{S}^2 =& \sum_{m^2} \ket{m^2-1} \bra{m^2}.
\end{align}

\subsection{Noncommutativity between translation and generator}
In order to define a noncommutative torus, we define $\hat{U}_3$ and $\hat{U}_4$ by
\begin{align}
    \hat{U}_3 =& \hat{S}^1, \\
    \hat{U}_4 =& \hat{S}^2 
\end{align}
Then, $\hat{U}_i$'s satisfy the following commutation relations:
\begin{align}
    \hat{U}_1\hat{U}_3 =& e^{-2\pi i \frac{a^1}{a^2}}\hat{U}_3\hat{U}_1, \\
    \hat{U}_2\hat{U}_4 =& e^{-2\pi i \frac{a^2}{a^1}}\hat{U}_4\hat{U}_2.
\end{align}
From these relations, we construct a noncommutative torus generated by $( \hat{U}_1 , \hat{U}_3)$ and  $(\hat{U}_2, \hat{U}_4)$.
The parameter $\theta$ of the corresponding NCT is
obtained from the commutation relation $\hat{U}_i \hat{U}_j = e^{-2\pi i \theta_{ij}} \hat{U}_j\hat{U}_i$ as
\begin{align}
    \theta_{13}
    =& 
    \frac{a^1}{a^2},
    \\
    \theta_{24}
    =& 
    \frac{a^2}{a^1}.
\end{align}
Hence, the effective Hamiltonian on each layer can be expressed as an operator defined on the noncommutative torus $\mathbb{T}_{\theta_{13}}$ and $\mathbb{T}_{\theta_{24}}$, and the total (effective) Hamiltonian of the bilayer system becomes an element of the algebra $\mathbb{T}_{\theta_{13}}\oplus\mathbb{T}_{\theta_{24}}$.
Thus the projector of the bilayer Hamiltonian is expressed as
\begin{align}
    \hat{P} = \hat{P}^1 \oplus \hat{P}^2,
\end{align}
where $\hat{P}$ is the projector in $\mathbb{T}_{\theta_{13}}\oplus\mathbb{T}_{\theta_{24}}$ and $\hat{P}^l$ is the projector in the layer $l$.
Here, it is worth noting that the relation between normalized trace of the bilayer system is not simply given by a summation of that of layer $l$ as $\tau(\hat{P}) = \tau^1(\hat{P}^1) + \tau^2(\hat{P}^2)$,
because the normalized trace $\tau^l$ is defined in a different subspace for each layer.
Therefore, to derive the GLT, we should examine the relation between $\tau^1$ and $\tau^2$.
Directly relating $\tau^1$ and $\tau^2$ is hard since the dimension of $\hat{U}_i$ is infinity as we have defined in Sec.~\ref{sec : Fourier expansion}.
To overcome this issue, we use the approximately finite (AF) algebra~\cite{kenneth1996c,brownc}.
This algebra defines the NCT as the inductive limit of the finite-dimensional matrix algebra, where we can relate the normalized traces for different layers more easily.

In the following, we follow the approach by Primsner and Voiculescu~\cite{10.2307/24714004} to relate the traces in different layers.
First, one represents $\theta$ with a continued fraction,
\begin{align}
    \theta = z_0 + \frac{1}{z_1 + 
    \frac{1}{z_2 + \frac{1}{z_3 + \frac{1}{\ddots}}}}
    =[z_0;z_1,\ldots ],
\end{align}
and then approximates $\theta$ with an approximant of $\theta$ defined as
\begin{align}
    \theta_n = z_0 + \frac{1}{\ddots + \frac{1}{z_n}}=\frac{p_n}{q_n},
\end{align}
where $p_n$ and $q_n$ are coprime integers.
When we approximate $\theta$ by $\theta_n$, we can represent the generators of the noncommutative torus with $q_n$-dimensional matrices and define IDoS using a trace of the finite-size matrices.
Next, to construct $\mathbb{T}^2_{\theta_{n+1}}$, we embed $\mathbb{T}^2_{\theta_n}$ and $\mathbb{T}^2_{\theta_{n-1}}$ to $q_{n+1}$-dimensional matrix with a suitable unitary transformation (for detail, see Eq.~(2) in Ref.~\cite{10.2307/24714004}).
Continuing this embedding, we define $\mathbb{T}^2_{\theta}$ as an inductive limit, $\mathbb{T}^2_{\theta} = \lim_{n\rightarrow\infty} \mathbb{T}^2_{\theta_n}$.
In this construction, the IDoS of $\mathbb{T}^2_{\theta}$ is also defined as the inductive limit from the IDoS of $\mathbb{T}_{\theta_n}$ which can be defined in the finite dimensional matrix algebra.
When we consider an approximant $\theta_n$ for the quasiperiodicity parameter $\theta = a^1/a^2$ for the bilayer system,
we effectively consider a superlattice made of $q_n$ sites in layer $1$ and $p_n$ sites in layer $2$ which we call
``the moir\'e unit cell''.
When we consider the system of the length $p_n$ with the periodic boundary condition, we may use a phase matrix $P_{\theta_n}$ and a clock matrix $C_{\theta_n}$ as an representation of $\hat{U}_1$ and $\hat{U}_3$:
\begin{align}
    \{ P_{ \theta_n=\frac{p_n}{q_n}} \}_{ij} =& \begin{cases}
        e^{2\pi i \theta_{n}j} & i=j \\
        0 & \mathrm{others}
    \end{cases},
    \\
    \{ C_{ \theta_n=\frac{p_n}{q_n}} \}_{ij} =& \begin{cases}
        1 & j-i = 1 \\
        1 & (i,j) = (q_n,1)  \\
        0 & \mathrm{others}
    \end{cases},
\end{align}
for $1 \leq i,j\leq q_n$.
Using these matrices, the representations of $\hat{U}$s under this approximation are
\begin{align}
    \hat{U}_1 =& P_{\theta_n}, \\
    \hat{U}_2 =& P_{\theta_n^{-1}}, \\
    \hat{U}_3 =& C_{\theta_n}, \\
    \hat{U}_4 =& C_{\theta_n^{-1}}.
\end{align}
The dimension of the $\hat{U}_1$ and $\hat{U}_3$ is $q_n$ and $\hat{U}_2$ and $\hat{U}_4$ is $p_n$.
Thus, in the finite-dimensional matrix algebra, the IDoS of the layer $1$ is expressed as
\begin{align}
    \mathrm{IDoS}^1(G) = \frac{1}{q_n} \mathrm{Tr}(\hat{P}(G)).
\end{align}
Since the Hamiltonian of bilayer system is the direct sum of $H^1_\text{eff}$ and $H^2_\text{eff}$,
the IDoS of the bilayer system and those of layer 1 and 2 satisfy the following relation: 
\begin{align}
    (p_n + q_n)\mathrm{IDoS}(G) = q_n \mathrm{IDoS}^1(G)
    +
    p_n \mathrm{IDoS}^2(G).
\end{align}
According to Ref.~\cite{rieffel1981c}, the IDoS for each layer is given by $\mathrm{IDoS}^1(G) = \mathfrak{n}_{1}^{1} + \theta_n \mathfrak{n}_{2}^{1}\quad (\mathfrak{n}_{1}^{1},\mathfrak{n}_{2}^{1} \in \mathbb{Z})$ and
$\mathrm{IDoS}^2(G) = \mathfrak{n}_{2}^{2} + \theta_n^{-1} \mathfrak{n}_{1}^{2} \quad (\mathfrak{n}_{1}^{2},\mathfrak{n}_{2}^{2} \in \mathbb{Z})$,
which leads to
\begin{align}
    \mathrm{IDoS}(G) = & (\mathfrak{n}_{1}^{1} + \mathfrak{n}_{1}^{2}) \frac{q_n}{(p_n + q_n)} +
    (\mathfrak{n}_{2}^{1} + \mathfrak{n}_{2}^{2}) \frac{p_n}{(p_n + q_n)}.
\end{align}
Taking the inductive limit of $p_n/ q_n \rightarrow \theta = a^1/ a^2$, we obtain
\begin{align}
    \mathrm{IDoS}(G) = & \frac{1}{\frac{1}{a^1} + \frac{1}{a^2}}\Big[ 
    \frac{\mathfrak{n}_1}{a^1}
    +
    \frac{\mathfrak{n}_2}{a^2}
    \Big]. 
\end{align}
Here, $\mathfrak{n}_{k} = \sum_{l=1,2} \mathfrak{n}_{k}^{l}$.
As a result, we can label $\mathrm{IDoS}$ of the bilayer system with two integers.
This result is consistent with the previous results including Fig.~6.3 in Ref.~\cite{PRODAN2019135}, and Eq.~(\ref{eq : GLT of 1D from SCN}) in Appendix \ref{sec : Sliding Chern number} from the SCN approach.

As one can easily see, $q_n$ is the number of unit cells inside the moir\'e unit cell of layer $1$ and $p_n$ is that of layer $2$.
Therefore, $q_n/(p_n+q_n)$ is the density of unit cells of layer $1$ inside the moir\'e unit cell. 
Thus relating the trace of each layer can be intuitively understood as introducing the ratio of the number of unit cells of two layers.
This allows us to extend the above formalism to multilayer and higher dimensional systems in a straight forward way.

In a similar manner, we can also extend GLT to the $N$-layer system.
Considering $N$-layer system with lattice constant $a^{l}$,
we can write IDoS of layer $l$ ($\mathrm{IDoS}^l$) as
\begin{align}
    \mathrm{IDoS}^l = \sum_{l^\prime = 1}^{N} \mathfrak{n}_{l^\prime}^{l}\frac{a^{l}}{a^{l^\prime}}.
\end{align}
Here, $\{ \mathfrak{n}_{1}^{l}, \ldots , \mathfrak{n}_{N}^{l} \}$ are integers.
From the relationship between $\mathrm{IDoS}^l$ and the IDoS of the entire system,
\begin{align}
    \left(\sum_{l=1}^{N} \frac{1}{a^{l}} \right)\mathrm{IDoS} 
    =& 
    \sum_{l} \frac{1}{a^{l}}\mathrm{IDoS}^l,
\end{align}
we obtain 
\begin{align}
    \mathrm{IDoS}
    =& 
    \frac{
    \left(
    \sum_{l = 1}^{N} \frac{\mathfrak{n}_l}{a^{l}}
    \right)
    }{
    \left(\sum_{l=1}^{N} \frac{1}{a^{l}} \right)
    }    \quad
    (\mathfrak{n}_{l^\prime} = \sum_{l} \mathfrak{n}_{l^\prime}^{l} ).
    \label{eq : IDoS of N layer}
\end{align}

\section{Gap labeling theorem of $D$-dimensional MLTFs}\label{sec : Gap labelling theorem of $D$-dimensional MLTFs}
In this section, we extend the GLT to $D$-dimensional MLTF heterostructures.

We first summarize our notations to describe $D$-dimensional MLTFs.
In $D$-dimensional MLTFs, we define $i$th primitive vector of layer $l$ as $\boldsymbol{a}^{l}_{i}$, and its reciprocal vector $\boldsymbol{b}^{l}_{i}$ is defined such that $\boldsymbol{a}^{l}_{i} \cdot \boldsymbol{b}^{l}_{j} = \delta_{ij}$.
Alternatively, we also denote $i$th primitive (reciprocal) vector in layer $l$ as $\boldsymbol{a}_{(l-1)D + i} = \boldsymbol{a}^{l}_{i}$.
In $D$-dimensional system, we can express $D$-dimensional volume of the region spanned by $\boldsymbol{a}_{i_1}, \cdots ,\boldsymbol{a}_{i_D}$ as $S_{i_1\ldots i_D} = |\det(\boldsymbol{a}_{i_1} , \cdots , \boldsymbol{a}_{i_D})|$.
When we replace $\boldsymbol{a}_{i_D}$ with unit vector $\boldsymbol{e}_{D}$ such that $\boldsymbol{e}_{D}\cdot \boldsymbol{a}_{i_j} = 0 \ (j = 1,\cdots D-1)$, we can omit $D$th index $i_D$ from $S$ and define
\begin{align}
    S_{i_1\ldots i_{D-1}} = |\det(\boldsymbol{a}_{i_1} , \cdots , \boldsymbol{a}_{i_{D-1}} , \boldsymbol{e}_{D})|.
\end{align}
In a similar manner, we can also define $S_{i_{1}\ldots i_{d}}$ as
\begin{align}
    S_{i_1\ldots i_{d}} = 
    |\det(\boldsymbol{a}_{i_1},
    \cdots ,
    \boldsymbol{a}_{i_{d}},
    \boldsymbol{e}_{d+1}, 
    \cdots ,
    \boldsymbol{e}_{D})|,
\end{align}
where $\boldsymbol{a}_{i}\cdot \boldsymbol{e}_{j} = 0 \ ( i = i_1,\ldots, i_d, j = d+1 ,\ldots , D)$.
In particular, if some vectors from $\boldsymbol{a}_{i_1}$ to $\boldsymbol{a}_{i_D}$ are parallel with each other, $S_{i_1 \ldots i_D} = 0$.
To label sites in layer $l$, we use integers $\{ n^{l}_{i}\}_{i=1,\ldots , D}$ and the parameter $\tilde{\boldsymbol{r}} \in \{ \sum_{d=1}^{D} \tilde{n}^{l}_{d} \boldsymbol{a}^{l}_{d} | 0 \leq \tilde{n}^{l}_{d} < 1 \}$ that specifies the origin of layer $l$ in the $\mathbb{R}^D$ space,
where the coordinate of the site $ \boldsymbol{r} \in \mathbb{R}^D$ is represented as
\begin{align}\label{eq : RD and ZD}
    \boldsymbol{r} = \sum_{d=1}^{D} n^{l}_{d} \boldsymbol{a}^{l}_{d}+\tilde{\boldsymbol{r}}.
\end{align}

\subsection{2D systems}
First, we extend Ref.~\cite{rosa2021topological} to two-dimensional heterobilayer systems.
Then, we derive GLT for the two-dimensional multilayer system.

\subsubsection{bilayer case}
We consider bilayer systems.
In this case, we have four primitive vectors and the parameter of the noncommutative torus for the layer $1$ is given by
\begin{align}\label{eq : noncommutative parameter layer 1 of 2D bilayer}
    \theta^{1}
    =
    \frac{1}{S_{34}}
    \begin{pmatrix}
    0 & 0 & S_{13} & S_{14} \\
    0 & 0 & S_{23} & S_{24} \\
    -S_{13} & -S_{23} & 0 & 0 \\
    -S_{14} & -S_{24} & 0 & 0
    \end{pmatrix}.
\end{align}
The derivation of $\theta^{1}$ is as follows.
In layer $1$ of the two-dimensional bilayer, the period of the quasiperiodic pattern is equal to the primitive vectors of layer $2$.
Specifically, omitting $\boldsymbol{\xi}^1(\boldsymbol{r})=\boldsymbol{0}$, we can represent the quasiperiodic coupling term only with $\boldsymbol{\xi}^2(\boldsymbol{r})$
which is the position of electron on layer $1$ in the unit cell of layer $2$.
Writing $\boldsymbol{\xi}^2(\boldsymbol{r}) = \tilde{\xi}_3(\boldsymbol{r})\boldsymbol{a}_3+\tilde{\xi}_4(\boldsymbol{r})\boldsymbol{a}_4$ with $\tilde{\xi}_j(\boldsymbol{r}) \in [0,1)$, 
we define
\begin{align}
    \hat{u}_{j} =& \sum_{ \boldsymbol{r}^1 \in \mathbb{R}^2}
    e^{2\pi i \boldsymbol{\xi}^2(\boldsymbol{r}^1)\cdot\boldsymbol{b}_j}\ket{\boldsymbol{r}^1}\bra{\boldsymbol{r}^1}
    =\sum_{ \boldsymbol{r}^1 \in \mathbb{R}^2}
    e^{2\pi i \tilde{\xi}_j(\boldsymbol{r}^1)}
    \ket{\boldsymbol{r}^1}\bra{\boldsymbol{r}^1}
    ,
\end{align}
where $j = 3,4$ and $\boldsymbol{r}^1$ runs the position of sites in layer $1$ in Eq.~(\ref{eq : RD and ZD}).
Denoting the shift operators that translate the layer $1$ by $-\boldsymbol{a}_1$ and $-\boldsymbol{a}_2$ as $\hat{u}_1$ and $\hat{u}_2$, we obtain
\begin{align}
    \hat{u}_{3} \hat{u}_{1} = \sum_{ \boldsymbol{r}^1 \in \mathbb{R}^2}
    e^{2\pi i 
    \boldsymbol{\xi}^2(\boldsymbol{r}^1-\boldsymbol{a}_1)\cdot\boldsymbol{b}_3
    }
    \ket{\boldsymbol{r}^1-\boldsymbol{a}_1}\bra{\boldsymbol{r}^1-\boldsymbol{a}_1},
\end{align}
and 
\begin{align}
    \hat{u}_{1} \hat{u}_{3}  = \sum_{ \boldsymbol{r}^1 \in \mathbb{R}^2}
    e^{2\pi i 
    \boldsymbol{\xi}^2(\boldsymbol{r}^1)\cdot\boldsymbol{b}_3
    }
    \ket{\boldsymbol{r}^1-\boldsymbol{a}_1}\bra{\boldsymbol{r}^1-\boldsymbol{a}_1}.
\end{align}
Hence the commutation relation of $\hat{u}_1$ and $\hat{u}_3$ is
\begin{align}
    \hat{u}_{1}\hat{u}_{3} = e^{2\pi i
    (
    \boldsymbol{\xi}^2(\boldsymbol{r}^1)
    -
    \boldsymbol{\xi}^2(\boldsymbol{r}^1-\boldsymbol{a}_1)
    )\cdot
    \boldsymbol{b}_3
    }\hat{u}_{3}\hat{u}_{1},
\end{align}
where this phase factor does not depend on $\boldsymbol{r}^1$.
Indeed, when we expand $\boldsymbol{a}_1$ as
\begin{align}
    \boldsymbol{a}_1 = \tilde{\theta}_{13}\boldsymbol{a}_3 + \tilde{\theta}_{14}\boldsymbol{a}_4, 
\end{align}
$\tilde\xi_3(\boldsymbol{r}^1-\boldsymbol{a}_1) \equiv \tilde\xi_3(\boldsymbol{r}^1) - \tilde{\theta}_{13} \mod{1}$ holds,
so that
\begin{align}
    e^{2\pi i (
    \boldsymbol{\xi}^2(\boldsymbol{r}^1)
    -
    \boldsymbol{\xi}^2(\boldsymbol{r}^1-\boldsymbol{a}_1)
    )\cdot
    \boldsymbol{b}_3
    } = e^{2\pi i \tilde{\theta}_{13}}.
\end{align}
From the definition of reciprocal vectors,
\begin{align}
    |\tilde{\theta}_{ij}| 
    =& |\boldsymbol{a}_i\cdot \boldsymbol{b}_j|
    = \frac{S_{i\bar{j}}}{S_{34}},
\end{align}
where $\{j,\bar{j}\}=\{3,4\}$.
As the sign does not affect the derivation of GLT, we may use $S_{i\bar{j}}/S_{34}$ instead of $\tilde{\theta}_{ij}$. Thus, defining
\begin{align}
    (\hat{U}_{1},\hat{U}_{2},\hat{U}_{3},\hat{U}_{4})
    =
    (\hat{u}_{1},\hat{u}_{2},\hat{u}_{4},\hat{u}_{3}),
\end{align}
we obtain the noncommutative parameters in Eq.~(\ref{eq : noncommutative parameter layer 1 of 2D bilayer}).
Applying Eq.~\eqref{eq: IDos G} for $\theta^{1}$, IDoS of the associated NC torus is given by
\begin{align}
    \mathrm{IDoS}^1 =& \mathfrak{n}_{\emptyset}^{1} + \sum_{J \subset \{1,2,3,4\}}\mathfrak{n}_{J}^{1} \mathrm{Pf}(\theta^{1}_J) \nonumber\\
    =& \mathfrak{n}^{1}_{\emptyset} + \sum_{i=1,2}\sum_{j=3,4} \mathfrak{n}_{ij}^{1} \frac{S_{ij}}{S_{34}}
    +
    \mathfrak{n}_{1234}^{1}\frac{S_{12}}{S_{34}},
\end{align}
where $\theta^{1}_{J}$ is a submatrix of $\theta^{1}$ defined in Eq.~(\ref{eq : def theta n}), and we used following formula
\begin{align}
    &S_{13}S_{24} - S_{23}S_{14} \nonumber\\
    &=
    (\boldsymbol{a}_1\times \boldsymbol{a}_3 ) \cdot (\boldsymbol{a}_2 \times \boldsymbol{a}_4 ) -(\boldsymbol{a}_2 \times \boldsymbol{a}_3 ) \cdot (\boldsymbol{a}_1 \times \boldsymbol{a}_4 ) \nonumber\\
    &= (\boldsymbol{a}_1\times \boldsymbol{a}_2 ) \cdot (\boldsymbol{a}_3 \times \boldsymbol{a}_4 )
    = S_{12}S_{34}.
\end{align}
This result means that we need six integers to label energy gaps of a single layer in two-dimensional bilayer system, which coincides with the result in Ref.~\cite{PhysRevB.104.035306}.

Similarly, the IDoS for layer $2$ is given by
\begin{align}
    \mathrm{IDoS}^2 = \mathfrak{n}_{\emptyset}^{2} + \sum_{i=1,2}\sum_{j=3,4} \mathfrak{n}_{ij}^{2} \frac{S_{ij}}{S_{12}}
    +
    \mathfrak{n}_{3412}^{2}\frac{S_{34}}{S_{12}}.
\end{align}
Next, we glue these two tori following the procedure we have discussed in the previous section.
As a result, we obtain IDoS of two-dimensional bilayer system as
\begin{align}
    \mathrm{IDoS}
    &= \frac{1}{S_{12} + S_{34}}
    \left[
    \sum_{i = 1}^{4} \sum_{j = i+1}^{4}
    \mathfrak{n}_{ij} S_{ij}
    \right].
\end{align}

\subsubsection{trilayer and $N$-layer cases}
First, we consider two-dimensional trilayer system.
In this case, we have six primitive vectors and the noncommutative parameter of layer $1$ is
\begin{align}
    \theta^{1}
    =
    \begin{pmatrix}
    0 & 0 & \frac{S_{13}}{S_{34}} & \frac{S_{14}}{S_{34}} & \frac{S_{15}}{S_{56}} & \frac{S_{16}}{S_{56}} \\
    0 & 0 & \frac{S_{23}}{S_{34}} & \frac{S_{24}}{S_{34}} & \frac{S_{25}}{S_{56}} & \frac{S_{26}}{S_{56}} \\
    -\frac{S_{13}}{S_{34}} & -\frac{S_{23}}{S_{34}} & 0 & 0 & 0 & 0 \\
    -\frac{S_{14}}{S_{34}} & -\frac{S_{24}}{S_{34}} & 0 & 0 & 0 & 0 \\
    -\frac{S_{15}}{S_{56}} & -\frac{S_{25}}{S_{56}} & 0 & 0 & 0 & 0 \\
    -\frac{S_{16}}{S_{56}} & -\frac{S_{26}}{S_{56}} & 0 & 0 & 0 & 0 
    \end{pmatrix}.
\end{align}
From Eq.~\eqref{eq: IDos G}, the IDoS of layer $1$ is
\begin{align}\label{eq : IDoS1 of two dimensional trilayer}
    \mathrm{IDoS}^{1}
    =&
    \mathfrak{n}_{\emptyset}^{1} + \sum_{i=1,2}\sum_{j = 3,4} \mathfrak{n}_{ij}^{1}\frac{S_{ij}}{S_{34}}
    + \sum_{i=1,2}\sum_{j = 5,6} \mathfrak{n}_{ij}^{1}\frac{S_{ij}}{S_{56}}
    \nonumber\\
    &
    +
    \frac{\mathfrak{n}_{1234}^{1} S_{12}}{S_{34}}
    +
    \frac{\mathfrak{n}_{1256}^{1} S_{12}}{S_{56}}
    +
    \sum_{i = 3,4,j=5,6} \mathfrak{n}_{12ij}^{1} \frac{S_{12}S_{ij}}{S_{34}S_{56}}.
\end{align}
In Eq.~(\ref{eq : IDoS1 of two dimensional trilayer}), the first five terms appear as the combination of $\mathrm{IDoS}^1$ in bilayer systems, while the last term does not. 
For example, fifth term comes from the bilayer between layer $1$ and layer $2$. In this case, quasiperiodic pattern generated by primitive vectors $\boldsymbol{a}_3$ and $\boldsymbol{a}_4$ opens the energy gaps.
In contrast, the last term with $\mathfrak{n}_{12ij}^{1}$ treats energy gap originates from the quasiperiodic pattern generated by primitive vectors $\boldsymbol{a}_i$ of layer $2$ and $\boldsymbol{a}_j$ of layer $3$, reflecting the trilayer nature.
Combining IDoS for each layer, we obtain
\begin{align}
    &
    \left(
    \frac{1}{S_{12}}
    +
    \frac{1}{S_{34}}
    +
    \frac{1}{S_{56}}
    \right)\mathrm{IDoS}
    \nonumber\\
    &=
    \frac{1}{S_{12}S_{34}S_{56}}
    \Big[
    (\mathfrak{n}_{\emptyset}^{1}
    +\mathfrak{n}_{3412}^{2}
    +\mathfrak{n}_{5612}^{3}
    )S_{34}S_{56}
    \nonumber \\
    &+ 
    (
    \mathfrak{n}_{1234}^{1}
    +\mathfrak{n}_{\emptyset}^{2}
    +\mathfrak{n}_{5634}^{3}
    )S_{12}S_{56} 
    \nonumber\\
    &+ 
    (
    \mathfrak{n}_{1256}^{1}
    +\mathfrak{n}_{3456}^{2}
    +\mathfrak{n}_{\emptyset}^{3}
    )
    S_{12}S_{34}
    \nonumber\\
    &+
    \sum_{i = 1,2} \sum_{j = 3,4} 
    (
    \mathfrak{n}_{ij}^{1}
    +\mathfrak{n}_{ij}^{2}
    +\mathfrak{n}_{56ij}^{3}
    )S_{ij} S_{56}
    \nonumber\\
    &+
    \sum_{i = 1,2} \sum_{j = 5,6} (
    \mathfrak{n}_{ij}^{1}
    +\mathfrak{n}_{34ij}^{2}
    +\mathfrak{n}_{ij}^{3}
    ) S_{ij}S_{34}
    \nonumber\\
    &+
    \sum_{i = 3,4} \sum_{j = 5,6} (
    \mathfrak{n}_{12ij}^{1}
    +\mathfrak{n}_{ij}^{2}
    +\mathfrak{n}_{ij}^{3}
    ) S_{ij}S_{12}
    \Big].
\end{align}
Since labels such as $\mathfrak{n}_{ij}^{1}+\mathfrak{n}_{ij}^{2}+\mathfrak{n}_{56ij}^{3}$ appear only as a combination in the expression for the IDoS,
we regard them as a single label.
Therefore, redefining labels, we obtain
\begin{align}
    \mathrm{IDoS} = \frac{
    \sum_{\substack{
    J \subset \{1,\ldots,6\} \\
    |J| = 2
    }} \mathfrak{n}_{J}\frac{1}{S_{J}}
    }{\frac{1}{S_{12}}+\frac{1}{S_{34}}+\frac{1}{S_{56}}},
\end{align}
where $J$ labels a set of two integers from $1$ to $6$.
The number of combination to choose two integers $J$ from $1$ to $6$ is 15.
Thus to label one-dimensional trilayer, we need 15 integers.
In a similar manner, to label two dimensional $N$-layer systems, 
we need to choose two vectors from $2N$ primitive vectors, which leads to ${_{2N}}C_{2} = N(2N-1)$ integers.

\subsection{General $D$-dimensional $N$-layer systems}
Similarly, we can also formulate the GLT in the $D$-dimensional $N$-layer MLTFs.
The noncommutative parameter of the NCT for layer 1 is written as
\begin{align}
    \theta^{1}
    =&
    \begin{pNiceMatrix}
    O & \Theta \\
    -^t\Theta & O
    \end{pNiceMatrix}, \nonumber\\
    \Theta
    =&
    \begin{pNiceMatrix}
     (\boldsymbol{a}_{1}\cdot\boldsymbol{b}_{D+1}) & \Cdots & (\boldsymbol{a}_{1}\cdot\boldsymbol{b}_{DN})\\
     \Vdots & \Ddots & \Vdots \\
     (\boldsymbol{a}_{D}\cdot\boldsymbol{b}_{D+1}) & \Cdots & (\boldsymbol{a}_{D}\cdot\boldsymbol{b}_{DN})
    \end{pNiceMatrix}.
\end{align}
From this matrix, we choose submatrix $\theta^{1}_{J}$ and calculate the Pfaffian.
When $\dim\theta^{1}_{J} = 2d$, a condition to obtain nonzero Pfaffian is to choose $d$ indices from $1,\ldots,D$ and $d$ indices from $D+1,\ldots, DN$. We denote chosen indices as $\mu \subset \{ 1,\ldots,D \}$ and $\nu \subset \{ D+1,\ldots,DN \}$.
Then, the Pfaffian of the submatrix is given by
\begin{align}
    \mathrm{Pf}(\theta^{1}_{J})
    =
    \det
    \begin{pNiceMatrix}
    (\boldsymbol{a}_{\mu_1} \cdot \boldsymbol{b}_{\nu_1}) & 
    \Cdots & 
    (\boldsymbol{a}_{\mu_1} \cdot \boldsymbol{b}_{\nu_d}) \\
    \Vdots & \Ddots & \Vdots \\
    (\boldsymbol{a}_{\mu_d} \cdot \boldsymbol{b}_{\nu_1}) & 
    \Cdots & 
    (\boldsymbol{a}_{\mu_d} \cdot \boldsymbol{b}_{\nu_d}) \\
    \end{pNiceMatrix},
\end{align}
where $J = \mu \cup \nu$.
When $d = D$, this simplifies to
\begin{align}
    \mathrm{Pf}(\theta^{1}_{J})
    =&
    \det
    \begin{pNiceMatrix}
    (\boldsymbol{a}_{\mu_1} \cdot \boldsymbol{b}_{\nu_1}) & 
    \Cdots & 
    (\boldsymbol{a}_{\mu_1} \cdot \boldsymbol{b}_{\nu_D}) \\
    \Vdots & \Ddots & \Vdots \\
    (\boldsymbol{a}_{\mu_D} \cdot \boldsymbol{b}_{\nu_1}) & 
    \Cdots & 
    (\boldsymbol{a}_{\mu_D} \cdot \boldsymbol{b}_{\nu_D}) \\
    \end{pNiceMatrix} \nonumber\\
    =&
    \det
    \begin{pNiceMatrix}
    - & \boldsymbol{a}_{\mu_1} & - \\
     & \Vdots & \\
    - & \boldsymbol{a}_{\mu_D} & - \\
    \end{pNiceMatrix}
    \det
    \begin{pNiceMatrix}
    \mid &  & \mid \\
    \boldsymbol{b}_{\nu_1} & \Cdots & \boldsymbol{b}_{\nu_D} \\
    \mid &  & \mid \\
    \end{pNiceMatrix}
    \nonumber\\
    =&
    \frac{S_{\mu}}{S_{\nu}} = \frac{S^{1}}{S_{\nu}}.
\end{align}
In the other cases, we obtain
\begin{align}
    &\mathrm{Pf}(\theta^{1}_{J})\nonumber\\
    &=
    \det\left[
    \begin{pNiceMatrix}
    - & \boldsymbol{a}_{\mu_1} & - \\
     & \Vdots & \\
    - & \boldsymbol{a}_{\mu_d} & - 
    \end{pNiceMatrix}
    \begin{pNiceMatrix}
    \mid &  & \mid \\
    \boldsymbol{b}_{\nu_1} & \Cdots & \boldsymbol{b}_{\nu_d} \\
    \mid &  & \mid 
    \end{pNiceMatrix}
    \right]
    \nonumber\\
    &=
    \det\left[
    \begin{pNiceMatrix}
    - & \boldsymbol{a}_{\bar{\mu}_1} & - \\
     & \Vdots & \\
    - & \boldsymbol{a}_{\bar{\mu}_{D-d}} & - \\
    - & \boldsymbol{a}_{\mu_1} & - \\
     & \Vdots & \\
    - & \boldsymbol{a}_{\mu_d} & - 
    \end{pNiceMatrix}
    \begin{pNiceMatrix}
    \mid &  & \mid &\mid &  & \mid \\
    \boldsymbol{b}_{\bar{\mu}_1} & \Cdots & \boldsymbol{b}_{\bar{\mu}_{D-d}}
    \boldsymbol{b}_{\nu_1} & \Cdots & \boldsymbol{b}_{\nu_d} \\
    \mid &  & \mid & \mid &  & \mid 
    \end{pNiceMatrix}
    \right]
    \nonumber\\
    &= (\pm 1) \frac{S^1}{
    S_{\bar{\mu}_1,\ldots,\bar{\mu}_{D-d},
    \nu_1,\ldots,\nu_d}
    }.
\end{align}
Here, $\mu \cup \bar{\mu} = \{1,\ldots D\}$,
and $\pm 1$ is the sign of the permutation of $\boldsymbol{a}_i$'s with $\{\bar{\mu}_1, \ldots,  \bar{\mu}_{D-d}, \mu_1, \ldots, \mu_d \}$.
This sign is not important to express IDoS, as we can get rid of it by redefining integer labels.
Using the results above, the IDoS for layer 1 is expressed as
\begin{align}
    \mathrm{IDoS}^1
    =
    \mathfrak{n}_{\emptyset}^{1}
    +
    \sum_{
    \substack{
    \mu \subset \{1,\ldots,D\} \\
    \nu \subset \{D+1,\ldots,DN\} \\
    |\mu \cup \nu| = D
    }
    }
    \mathfrak{n}_{\mu\nu}^{1} \frac{S^1}{S_{\mu\nu}}.
\end{align}
In a similar manner, we can also calculate the IDoS for layer $l$.
Using the relationship,
\begin{align}
    \left(\sum_{l}\frac{1}{S^l}\right)\mathrm{IDoS} = \sum_{l} \frac{1}{S^l}\mathrm{IDoS}^l ,
\end{align}
we obtain the IDoS for the $D$-dimensional $N$-layer as
\begin{align}\label{eq : GLT for D-dim N-layer}
    \mathrm{IDoS} =
    \frac{
    \sum_{\substack{
    J \subset \{1,\ldots,DN \} \\
    |J| = D
    }} \mathfrak{n}_{J}\frac{1}{S_{J}}
    }{\sum_{l}\frac{1}{S^l}}.
\end{align}
This formula shows that the number of choices of $J$ that has nonzero contribution is ${_{DN}}C_{D}$, which indicates that we need ${_{DN}}C_{D}$ integers to label the energy gap structure of $D$-dimensional $N$-layer MLTFs generally.

\section{Numerical calculation}\label{sec : Numerical calculation}
We demonstrate the validity of the obtained gap labels in the case of one-dimensional systems.
First, we consider the bilayer system.
We then move on to the trilayer system with large gaps between layers, which is ideal to project out the other layers into potentials.

\begin{figure}
    \centering
    \includegraphics[width = \linewidth]{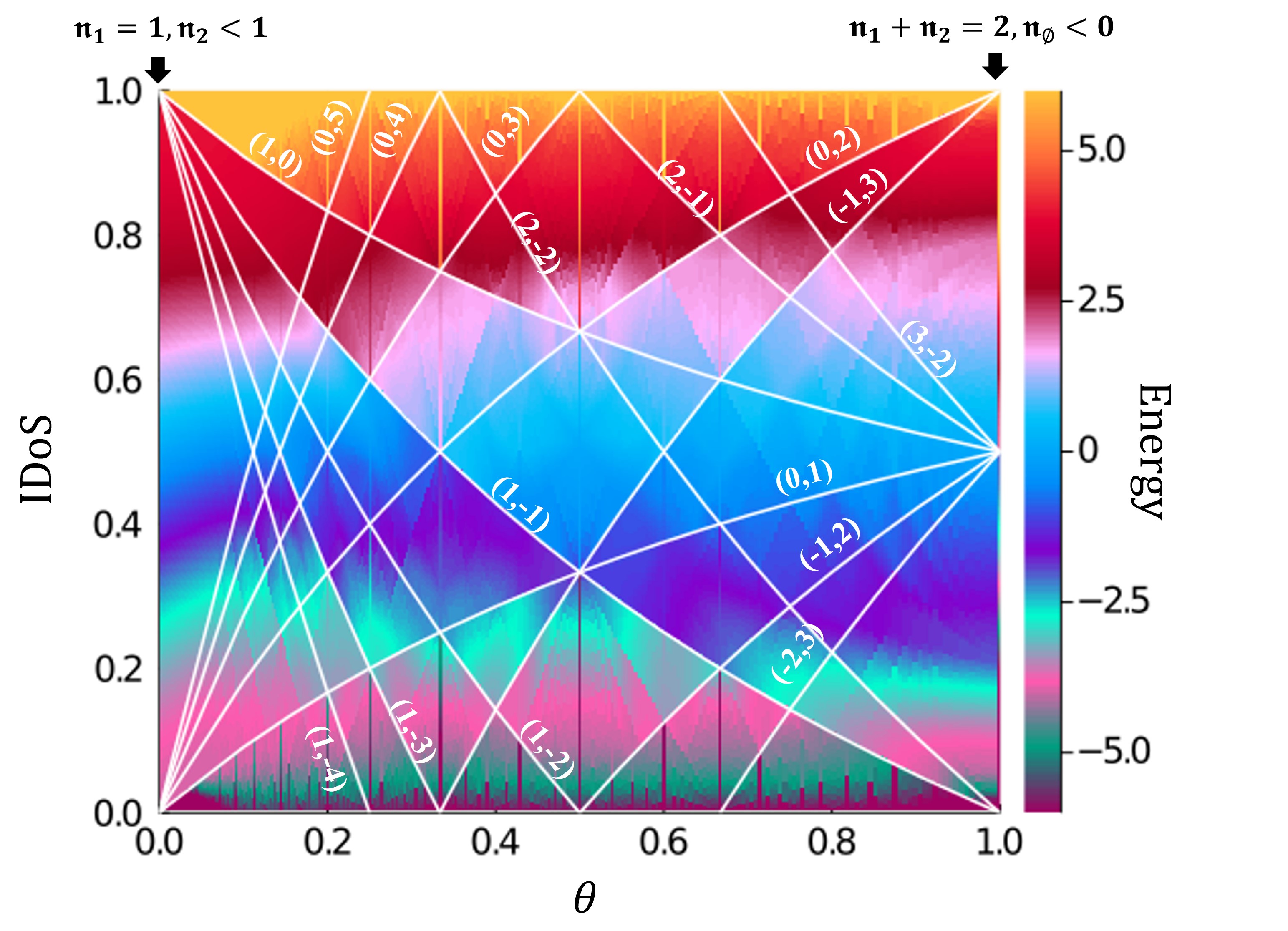}
    \caption{
        The integrated density of states (IDoS) of the bilayer system against lattice constant $\theta$.
        The colour changes with energy. The region where the colour changes sharply corresponds to the energy gap.
        White lines are energy gaps predicted by the GLT with labels $(\mathfrak{n}_1,\mathfrak{n}_2)$ defined in Eq.~(\ref{eq : IDoS of N layer}).
        The labels of energy gaps whose IDoS are $1$ at $\theta =0$ satisfies $\mathfrak{n}_1 = 1$ and $\mathfrak{n}_2 < 1$.
        The labels of energy gaps whose IDoS are $1$ at $\theta =1$ satisfies $\mathfrak{n}_1 + \mathfrak{n}_2 = 2$ and $\mathfrak{n}_2 < 0$.
    }
    \label{fig : IDoS 1D bilayer}
\end{figure}

First, we consider a bilayer consisting of layers with lattice constant $1$ and $\theta$.
The Hamiltonian is composed of the two single layer Hamiltonians with nearest neighbour coupling and an interlayer coupling term: 
\begin{align}
    \hat{H} =& \sum_{l=1,2}\sum_{n^l}
    \Big[  t^l \ket{n^l+1}\bra{n^l} + h.c. \Big]
    \nonumber\\
    &+
    \sum_{n^1,n^2} \Big[
    Ve^{-\gamma r(n^1,n^2)} \ket{n^1}\bra{n^2}
    +
    h.c.
    \Big].
\end{align}
Here $r(n^1,n^2)$ is the distance between site $n^1$ and $n^2$ which we define $r(n^1,n^2)=|n^1a^1-n^2a^2|$, and we take $t^1 = 1, t^2 = 2, V = 5, \gamma = 10$.
The result is shown in Fig.~\ref{fig : IDoS 1D bilayer}.
The color corresponds to the energy of eigenstates. The sharp color changes correspond to the energy gaps.
The white solid curves are the position of the energy gap predicted from the generalized GLT in Eq.~(\ref{eq : IDoS of N layer}), where we indicate the associated integer labels.
The white curves coincides with the position of the energy gap (where the color changes sharply) and shows that the generalized GLT holds in the present system.
Vertical lines without color change appears in fractional $\theta$s.
While we show the curves associated with a few labels, we can also reproduce positions of other energy gaps from the GLT as well.

Next, we show numerical calculations in a trilayer system.
Lattice constants are $a^1 = 1, a^2 = \alpha , a^3 = \alpha\beta $, and $\beta$ is fixed to $12/13$. The Hamiltonian is
\begin{align}
    \hat{H} =& \sum_{l=1}^{3}\sum_{n^l}
    \Big[
    t^l \big(\ket{n^l+1}\bra{n^l} + h.c \big)
    +
    E^l\ket{n^l}\bra{n^l}
    \Big]
    \nonumber\\
    &+
    \sum_{n^1,n^2} \Big[
    \left(Ve^{-\gamma r(n^1,n^2)}\right)^2 \ket{n^1}\bra{n^2}
    +
    h.c.
    \Big]
    \nonumber\\
    &+
    \sum_{n^2,n^3} \Big[
    \left(Ve^{-\gamma r(n^2,n^3)}\right)^2 \ket{n^2}\bra{n^3}
    +
    h.c.
    \Big]
    \nonumber\\
    &+
    \sum_{n^3,n^1} \Big[
    Ve^{-\gamma r(n^3,n^1)} \ket{n^3}\bra{n^1}
    +
    h.c.
    \Big]
    .
\end{align}
Here we take $t^1 = t^2 = t^3 = 2, V = 2 , \xi = 2$.
$E^l$'s are additional parameter to show energy gap clearly, and set $E^1 = 10, E^2 = 0, E^3 = -10$.

\begin{figure}[t]
    \centering
    \includegraphics[width = \linewidth]{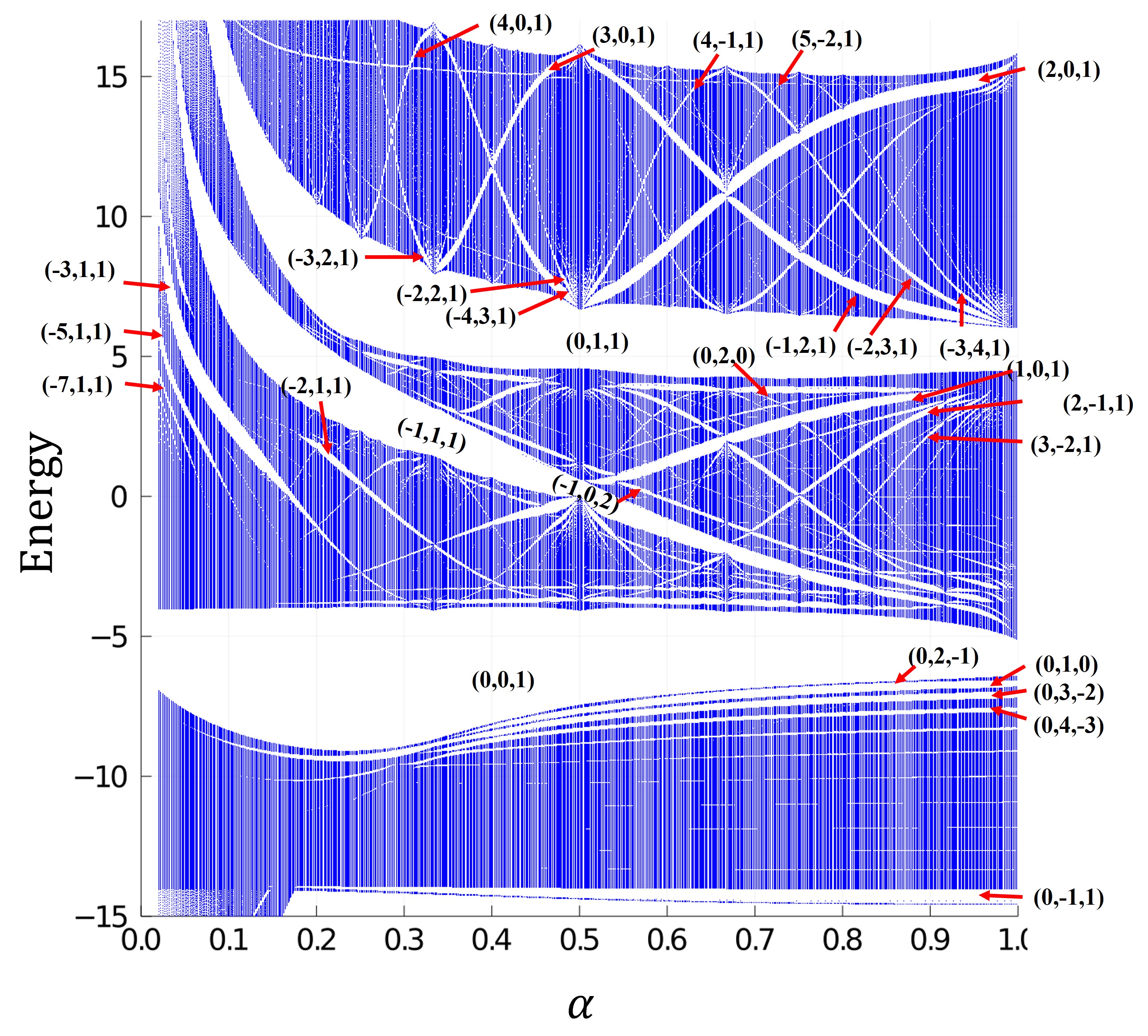}
    \caption{
    Trilayer model with large gaps. Here, we set $\beta = 12/13$.
    Parameters are $t_1 = t_2 = t_3 = 2, V = 2 , \xi = 2, E_1 = 10, E_2 = 0, E_3 = -10$.
    Blue points are the energy spectrum and red lines are the energy gap predicted by the GLT.
    Arrows indicate energy gap and their labels $(\mathfrak{n}_1,\mathfrak{n}_2,\mathfrak{n}_3)$.
    }
    \label{fig : trilayer large}
\end{figure}

The energy spectrum under the PBC is shown in Fig.~\ref{fig : trilayer large}.
Blue points are the energy spectrum and red arrows show the energy gap predicted by the GLT.
Among energy gaps, there are two trivial gaps $(\mathfrak{n}_1,\mathfrak{n}_2,\mathfrak{n}_3) = (0,0,1)$ and $(0,1,1)$.
$(0,0,1)$ corresponds to filling layer 3 only, and $(0,1,1)$ corresponds to filling layer 2 and 3.
These gaps open without quasiperiodicity, so stable against the change of $\alpha$.
In Fig.~\ref{fig : trilayer large}, the size of energy gaps are different.
This is due to the exponential decay of the interlayer coupling.
When the interlayer coupling takes large value, on the Bragg lines, the energy gap is also large.
Even in the case where the energy gaps are small, e.g. $(5,-2,1)$, we can also label energy gaps.
Therefore, also in the trilayer systems, we can predict energy gaps using GLT.

We note the limitation for the choices of $E^l$.
In this calculation, we added energy offsets to separate the energy bands of the three layers.
This allows us to focus on intraband energy gaps which can be captured by the GLT.
When energy bands are energetically close and hybridized with each other, interband energy gaps appear which are not necessarily captured by the GLT.
Characterization of those interband energy gaps remains a future problem.

\section{Discussions}
In this paper, we have extended the GLT to MLTFs using the approach from the noncommutative torus, which allows us to treat all layers on an equal footing. 
We derived the GLT for general $D$-dimensional $N$-layer systems by combining previous results on the GLT \cite{10.2307/24713853,rosa2021topological} and redefinition of the normalized trace for each layer 
using a continued fraction and  the inductive limit.
As a result, we have obtained the general expression of IDoS in Eq.~(\ref{eq : GLT for D-dim N-layer}), and found that the number of gap labels is generally given by $_{DN}C_D$.
In addition, the obtained GLT formula was found to be consistent with the result from the SCN approach. 
The present NCT approach also gives a reinterpretation of the result obtained from the SCN approach. (For details, see Appendix.~\ref{sec : Sliding Chern number}).

We note on the validity of treating the interlayer coupling as an effective quasiperiodic intralayer coupling for each layer.
In this paper, we focused on the situation where we can treat the interlayer coupling as an effective intralayer coupling for each layer.
This is a good approximation when the energy bands from different layers are energetically separated.
When the states from different layers are energetically close and hybridized due to the interlayer coupling, the resulting energy gaps do not arise from the quasiperiodicity and are not expected to be characterized by the GLT.
To be more precise, it is not evident whether one can rewrite the Hamiltonian with general interlayer couplings as an operator belonging to a NC torus without projecting to the effective intralayer coupling.

An interesting future direction is studying the relationship between the GLT and the flat bands in the MLTFs.
In MLTFs such as twisted bilayer graphene, an emergence of flat bands at certain angles has been reported.
This is a consequence of the gap opening between replicas of the energy band and originates from the quasiperiodicity.
Recently, study of the flat band in twisted bilayer graphene has been studied using Jacobi theta functions \cite{PhysRevLett.122.106405,PhysRevB.106.085140,PhysRevResearch.2.023237},
where one can define a commutation relationship for quasiperiodic functions in a similar manner to those for $S$ and $U$ used in Sec.~\ref{sec : Gap labelling theorem of one-dimensional multilayer thin films} \cite{mumford2013tata}.
One can interpret that the theta functions are describing the wave functions of quasiperiodic systems, while the NC torus is describing the Hamiltonian.
Thus, the GLT may also be useful to study the flat bands in quasiperiodic systems if a relationship between the GLT and the theta function is established.

\begin{acknowledgements}
We thank Mikio Furuta, Hosho Katsura and members in Koshino group in Osaka University for fruitful discussions. 
This work was supported by
JST CREST (JPMJCR19T3) (SK, TM), and JST PRESTO (JPMJPR19L9) (TM). 
MY was supported by Forefront physics and mathematics program to drive transformation (FoPM).
\end{acknowledgements}

\appendix

\section{Sliding Chern number}\label{sec : Sliding Chern number}
In MLTFs, a new topological index which comes from the sliding of two-dimensional layers is proposed independently in Refs.~\cite{PhysRevB.101.041112,PhysRevB.101.041113,PhysRevB.101.041410}.
Here, we call it the sliding Chern number (SCN) following Ref.~\cite{PhysRevB.101.041112}.
\subsection{Gedanken experiment to relate SCN and gap labels}
In Ref.~\cite{PhysRevB.101.041112}, a gedanken experiment was conducted to clarify the relationship between the SCN and a quantized charge transport.
This argument can be used to obtain GLT for one-dimensional bilayer systems with an extension.

We consider a one-dimensional bilayer system without the internal degree of freedom.
The lattice constant of the bottom layer is 1 and the top layer is $p/q$.
Here, $p$ and $q$ are coprime to each other, and two layers are connected through interlayer coupling.

In this case, we can define the moir\'e unit cell for the bilayer system and its lattice constant to be $q$.

When we slide the top layer by $q$, the system before and after the sliding is identical, yet electrons on the top layer are transferred.
With the constraint that the system must be identical in two cases,
the number of transferred electrons is a multiple of sites inside the moir\'e unit cell, $\mathfrak{n}_tq \ (\mathfrak{n}_t \in \mathbb{Z})$.

In a similar manner, we can also slide the bottom layer by $q$ into the opposite direction.
In this case, the number of transferred electrons is a multiple of $\mathfrak{n}_bp \ (\mathfrak{n}_b \in \mathbb{Z})$.

Since two slid bilayers are related through the translation of the entire system by $q$.
We can obtain the following equation 
\begin{align}
    \mathfrak{n}_t q = \mathfrak{n}_b p + N.
\end{align}
Here, $N$ is the number of electrons inside the moir\'e unit cell. In Ref.~\cite{PhysRevB.101.041112}, it was shown that $\mathfrak{n}_t$ and $\mathfrak{n}_b$ are the sliding Chern number.

To see the relationship of the above argument with the IDoS, we divide both sides with the system size $p+q$, which leads to
\begin{align}
    \mathrm{IDoS} = \mathfrak{n}_t \frac{q}{p+q} -\mathfrak{n}_b \frac{p}{p+q}.
\end{align}
From the definition, IDoS is the number of electrons divided by the system size $N/(p+q)$.
Finally, we take the incommensurate limit.
Replacing $p/q$ with an irrational number $\theta$, we obtain
\begin{align}\label{eq : GLT of 1D from SCN}
    \mathrm{IDoS} 
    = \frac{\mathfrak{n}_t - \mathfrak{n}_b \theta}{1+\theta}.
\end{align}
Therefore, we can also relate the SCN with IDoS.
This gedanken experiment does not depend on the detail of the Hamiltonian, and Eq.~(\ref{eq : GLT of 1D from SCN}) coincides with the result in Sec.~\ref{sec : Gap labelling theorem of one-dimensional multilayer thin films}.

\subsection{Two and higher dimensions}
In Ref. \cite{PhysRevB.104.035306}, the relationship between the energy gaps of hBN/Graphene/hBN trilayer heterostructure and integer labels were pointed out, and later the authors studied the relationship between the energy gap structure and the integer labels in general two-dimensional multilayer systems in Ref.~\cite{PhysRevResearch.4.013028}.
From these studies, they derived the following expression for the IDoS
\begin{align}\label{eq : GLT for 2D Koshino}
    \mathrm{IDoS}(G) = \mathfrak{n}_{\emptyset} + \sum_{i=1,2}\sum_{j=3,4}
    \mathfrak{n}_{ij} \frac{S_{ij}}{S_{12}}
    +\mathfrak{n}_{1234}\frac{S_{34}}{S_{12}},
\end{align}
which is consistent with the generalized GLT formula for $D=2, N=2$ case. 
More recently, Yamamoto and Koshino have pointed out that one can label energy gaps of the three-dimensional system under incommensurate potential with the third Chern number \cite{PhysRevB.105.115410}. This is consistent with our result of the generalized GLT, in that we can transform the Hamiltonian of $D$-dimensional lattice under incommensurate potential to NC $2D$-torus and the top Chern number is the $D$-th Chern number \cite{prodan2013non}.

\bibliography{ref}

\end{document}